\documentclass[twocolumn,preprintnumbers,amsmath,amssymb,amsfonts,prb,floatfix]{revtex4}

\usepackage{graphicx} 
\usepackage{dcolumn} 
\usepackage{bm} 

\begin{document}

\title{Inelastic scattering effects and the Hall resistance in a 4-probe ring}  

\author{G.~Metalidis}
  \email{georgo@mpi-halle.de}
\author{P.~Bruno}
  \email{bruno@mpi-halle.de}
\affiliation{%
Max-Planck-Institut f\"{u}r Mikrostrukturphysik, Weinberg 2,
D-06120 Halle, Germany} \homepage{http://www.mpi-halle.de}
\date{\today}

\begin{abstract}
Phase randomizing processes in mesoscopic systems can be described
in a phenomenological way within the Landauer-B\"{u}ttiker
formalism by attaching extra voltage probes to the sample. In this
paper, it is shown that a perturbation treatment of this idea
allows for the incorporation of such effects without the need of
giving up the efficiency of recursive techniques commonly used for
calculating the transmission coefficients. The technique is
applied to a 4-probe ring, where a Hall effect can be observed
that originates from quantum interference rather than a Lorentz
force acting on the electrons. The influence of inelastic
scattering on both the Hall resistance and the Aharonov-Bohm
oscillations in the longitudinal resistance are examined.
\end{abstract}

\maketitle

\section{Introduction}
Quantum coherence plays a prominent role in the transport
properties of mesoscopic systems; interference between different
electron trajectories can lead to interesting effects like weak
localization~\cite{weakloc} and Aharonov-Bohm (AB) oscillations.
Inelastic scattering events will destroy phase coherence and as
such interference effects will be smeared out. Modelling inelastic
scattering proves to be difficult because it is in general a
many-body problem. However, a proposal has been put forward some
years ago for incorporating these effects phenomenologically
within the Landauer-B\"{u}ttiker (LB) formalism (which is a single
particle theory), by attaching extra voltage probes to the
sample~\cite{Buettiker1}. Nevertheless, only few
papers~\cite{Damato, prevpapers} have made use of it because the
widely used standard recursive Green's function method is unable
to calculate the transmission coefficients between the extra
voltage probes. As such, one should in principle invert the
complete Hamiltonian to solve for all transmission coefficients
which is of course computationally very inefficient for large
systems. In this paper it will be shown how to treat the regime of
weak inelastic scattering with a perturbation approach to the
original voltage probe model, still keeping the computational
effort the same as needed for the standard recursive Green's
function method.

Our method is applied to a ring setup with 4 probes and a flux
piercing through like depicted in Fig.~\ref{Fig1}. In such a setup
a Hall voltage can be observed which is completely due to
interference of electrons travelling in opposite directions along
the ring, and which in principle does not rely on a Lorentz force
acting on the electrons~\cite{Robinson, Gartner}. In the current
paper, numerical results for a realistic two-dimensional (2D) ring
will be shown, taking into account inelastic scattering. The
observed Hall effect disappears when decreasing the phase
coherence length, showing again that the effect is solely due to
quantum interference.

\section{Modeling Inelastic Effects}
For calculating resistances in our system, we will use the LB
formalism which relates resistances to transmission probabilities
between the leads. In the formalism, the leads are thought to be
connected to large reservoirs with a well-defined chemical
potential and temperature. The currents through the leads and the
voltages on the reservoirs are related through (at temperature
$T=0$):
\begin{equation} \label{LBformalism}
I_p = \frac{2e^2}{h} \sum_{q} T_{pq} \, (V_p - V_q),
\end{equation}
where $p,q$ label the leads/reservoirs and $T_{pq}$ is the
transmission probability from lead $q$ to lead $p$. Although this
formalism is only valid when phase coherence is present in the
sample and the leads, phase breaking events must be taking place
in the reservoirs in order for them to have a well-defined
equilibrium distribution with a certain chemical potential. It is
this insight that was used by B\"{u}ttiker to arrive at the idea
that an extra lead connected to a reservoir can introduce a phase
breaking event~\cite{Buettiker1}. Indeed, if the current through
such a lead is fixed to be zero (this is what we call a voltage
probe), then for every electron that enters the lead and is
absorbed by the reservoir, another one has to come out. But since
equilibration is taking place in the reservoir, the electron
coming out is not coherent with the one going in. In this way, it
is possible to model inelastic effects in a phenomenological way.

In our calculations, we will attach a one-dimensional (1D) lead on
every site of our tight binding model in order to have a
homogeneous distribution of inelastic scattering centers. These
leads can be thought to extend in a direction perpendicular to the
2D sample (see Fig.~\ref{Fig1}). An adequate choice of the
potential in the 1D leads makes it possible that the influence of
these leads can be described by just adding a complex onsite
energy $- \mathrm{i} \eta$ to every site in the
lattice~\cite{Damato}. The parameter $\eta$ is controlled by the
hopping parameter between the 1D leads and the sample, and is
related to the inelastic scattering time as $\eta = \hbar / 2
\tau_\varphi$. However, the approach goes further than just adding
an imaginary potential; in order to conserve the total current in
the system, one has to solve Eq.~(\ref{LBformalism}) so that the
current through the extra 1D leads is zero. Before giving
expressions for the currents through the leads, we will introduce
some notation: the 1D voltage probes attached to the system to
simulate phase randomizing processes are labelled by a Greek
index, while the real physical leads connected to the sample
(referred to as contacts from now on) will be labelled by letters
$m,n,\ldots$. Now, by putting $I_\alpha = 0$ for all $\alpha$, and
solving for the voltages $V_\alpha$ on the voltage probes in terms
of the voltage differences $V_n - V_m$ between the contacts, one
can obtain an expression for the currents $I_n$ through the
contacts similar to Eq.~(\ref{LBformalism}), but now with
effective transmission probabilities incorporating the effect of
the voltage probes:
\begin{equation}
I_n = \frac{2e^2}{h} \sum_{m} T_{nm}^\text{eff} \, (V_n - V_m),
\end{equation}
where the effective transmission probabilities can be written as
follows:
\begin{equation} \label{efftrans}
T_{nm}^\text{eff} = T_{nm} + \sum_{\alpha} \frac{T_{n \alpha}
T_{\alpha m}} {S_{\alpha}} + \sum_{\alpha} \sum_{\beta \neq
\alpha} \frac{T_{n \alpha} T_{\alpha \beta} T_{\beta m}}
{S_{\alpha} S_{\beta}} + \cdots,
\end{equation}
with $S_{\gamma} = \sum_{l} T_{ \gamma l} + \sum_{\delta \neq
\gamma} T_{ \gamma \delta}$. This expression has a clear physical
interpretation: the first term describes transmission from contact
$m$ to contact $n$ without any inelastic process, the next term
incorporates a single scattering event, then double scattering and
so on.

A standard method commonly employed (because of its efficiency)
for calculating the transmission coefficients is the recursive
Green's function technique (reviewed in Ref.~\onlinecite{Ferry}).
However, this technique can give only access to the transmission
coefficients between the contacts $T_{nm}$. For calculating the
effective transmission coefficients in Eq.~(\ref{efftrans}), one
would also need the transmission probabilities between the
contacts and the voltage probes ($T_{n \alpha}$ and $T_{\alpha
m}$), and between the voltage probes themselves ($T_{\alpha
\beta}$). These are unavailable with the standard technique,
because one would need Green's functions connecting points at the
edges of the sample (where the contacts are attached) with points
inside the sample, whereas the technique only gives us Green's
functions between the left and right edge.

However, one could neglect all terms in Eq.~(\ref{efftrans})
involving two and more scattering events (by putting $T_{\alpha
\beta} = 0$), and keep only the direct transmission term together
with the single scattering term. This approximation is valid when
the phase coherence length of the sample is larger than the system
size. Although the standard technique still cannot give access to
$T_{n \alpha}$ needed for the second order term, a recursive
technique has been developed recently that allows to calculate
these with the same numerical effort needed for the standard
recursive Green's function method. Details of this technique can
be found in Ref.~\onlinecite{ourpaper}. In that way we are able to
treat the effective transmission coefficients to second order in
the phase randomization processes.

It should be noted that by neglecting higher order terms in
Eq.~(\ref{efftrans}), one will break current conservation rules
whenever a magnetic flux is present in the system; but in the weak
scattering regime, this error will be small. In all calculations
we present, the sum of currents flowing through leads 1 to 4
divided by the sum of their absolute values was smaller than
$10^{-4}$.

\section{Results for a 4-probe ring}
\begin{figure}
\includegraphics[width = 8cm]{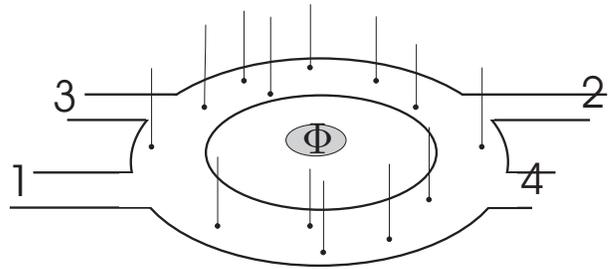}
\caption{ \label{Fig1} Schematical view of the 4-probe ring setup.
Inelastic scatterers will be modelled by attaching a 1D lead at
every scattering center. In real calculations, every site of the
tight binding model is attached to a voltage probe in order to
have homogeneous scattering.}
\end{figure}
We will consider a 4-probe ring in a two-dimensional electron gas,
like shown schematically on Fig.~\ref{Fig1}. All leads are
arranged to the left and right of the ring for computational
convenience. The Hamiltonian of the system can be written as:
\begin{equation}
H = \frac{1}{2 m^\star} \left ( \mathbf{p} - e \mathbf{A} \right
)^2 + V_\text{imp} + V_c,
\end{equation}
where $m^\star$ and $e$ are respectively the effective mass and
charge of the electron, $\mathbf{A}$ is the vector potential
created by the flux through the ring, $V_c$ is the confinement
potential defining the ring, and $V_\text{imp}$ is the elastic
impurity potential. By discretizing this Hamiltonian on a square
lattice with lattice parameter $a$, one obtains a tight binding
model:
\begin{eqnarray*}
H &=& \sum_{n} \sum_{m} \epsilon_{mn} |m,n\rangle \langle m,n| + \\
& \ & \left( t_{m,n}^{x} |m,n+1\rangle \langle m,n| + t_{m,n}^{y}
|m+1,n\rangle \langle m,n | + h.c. \right),
\end{eqnarray*}
with $(m,n)$ labelling the sites on the lattice, and
$\epsilon_{mn}$ the on-site energies. The hopping parameters are
given by:
\begin{equation}
t_{mn}^{x(y)} = -t \, \mathrm{e}^{-\mathrm{i} \, e / \hbar \int
\mathbf{A} \centerdot \mathrm{d} \mathbf{l}},
\end{equation}
with the integral evaluated along the hopping path and $t =
\hbar^2 / 2 m^\star a^2$.

The ring parameters were chosen in accordance to typical rings
fabricated experimentally in a 2DEG at the interface of an
GaAs-AlGaAs heterostructure. The density of the 2DEG was chosen to
be $n_s = 4 \times 10^{11} \, \text{cm}^{-2}$, corresponding to a
Fermi wavelength of $40 \, \text{nm}$. The ring has a mean radius
of $0.6 \, \mu \text{m}$, and the width of the arms is $200 \,
\text{nm}$, so that 10 channels are available for conduction. The
mobility of the electron gas is $\mu = 5 \times 10^5 \,
\text{cm}^2 \, \text{V}^{-1} \, \text{s}^{-1}$, giving a mean free
path of $5.2 \, \mu \text{m}$.
\begin{figure}
\includegraphics[width = 8cm]{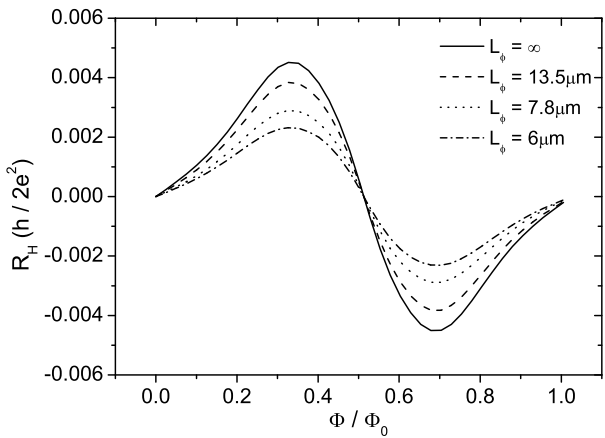}
\caption{ \label{Fig2} The Hall resistance $R_H = 1/2 \,
(R_{12,34} - R_{34,12})$  for different values of the phase
coherence length.}
\end{figure}

This translates into the following tight binding parameters. The
lattice parameter was chosen to be $a = 6.7 \, \text{nm}$, so that
$\lambda_F = 6\, a$ and the Fermi energy $E_f = 1.1 \, t$. The
mean radius of the ring is $89 \, a$, and the width $29 \, a$.
Elastic scattering was modelled with on-site energies distributed
randomly in the interval $[-0.127 \, t, 0.127 \,t]$, which gives a
mean free path of $l_m = 780 \, a$ (estimated in Born
approximation).

For this ring, we will calculate both the longitudinal resistance
$R_{12,12}$ and the transverse resistances $R_{12, 34}$ and
$R_{34, 12}$ (see Fig.~\ref{Fig1}), and the influence of inelastic
processes on the results will be made clear. We use the common
notation $R_{kl,mn} = (V_m - V_n)/I_k$ for a measurement where
current is supplied through contacts $k$ and $l$, and the voltage
difference $V_m - V_n$ is measured, fixing $I_m = I_n = 0$. In
terms of transmission coefficients, these resistances are given
by~\cite{Buettiker2}:
\begin{equation}
R_{kl, mn} = \frac{T_{mk} T_{nl} - T_{nk} T_{ml}}{D},
\end{equation}
where $D$ is a $3 \times 3$ subdeterminant of the matrix $A$
relating the currents through the four contacts to their voltages
[$I_m = A_{mn} V_n$, c.f.
Eq.~(\ref{LBformalism})]~\cite{Buettiker2}. $D$ is independent of
the indices $klmn$. When inelastic processes are included, the
transmission coefficients in these expressions have to be replaced
with the effective probabilities given in Eq.~(\ref{efftrans}).

Let's look at the resistance $R_{12,34}$. Suppose an electron
enters the ring through lead 1. It can reach lead 3 by different
paths: there is a direct path between lead 1 and lead 3, but there
is also a trajectory going like $1 \rightarrow 4 \rightarrow 2
\rightarrow 3$. These two trajectories will interfere with each
other, and fix the voltage on lead 3 (neglecting paths circling
the ring more than once). The same can be applied to lead 4: again
there is a direct and indirect path interfering and setting the
voltage on lead 4. It is clear that because of symmetry, the
voltage on lead 3 and lead 4 will be the same when no flux is
present through the ring. However, when a flux pierces through the
ring, time reversal symmetry is broken and the phase difference
between both paths going from 1 to 3 is different from that for
the paths going from 1 to 4, so that a voltage difference $V_3 -
V_4$ will develop. For a one-dimensional ring, this effect is
described in some detail in Ref.~\onlinecite{Robinson}, and
numerical calculations for a simple model are shown in
Ref.~\onlinecite{Gartner}. The effect is purely based on
interference; a Hall voltage will also be present when the
magnetic field is limited to the inside of the ring, and no flux
is going through the arms of the ring. Like in the original
Aharonov-Bohm proposal, the electron can feel the vector
potential, and is not subjected to a real Lorentz force.
\begin{figure}
\includegraphics[width = 8cm]{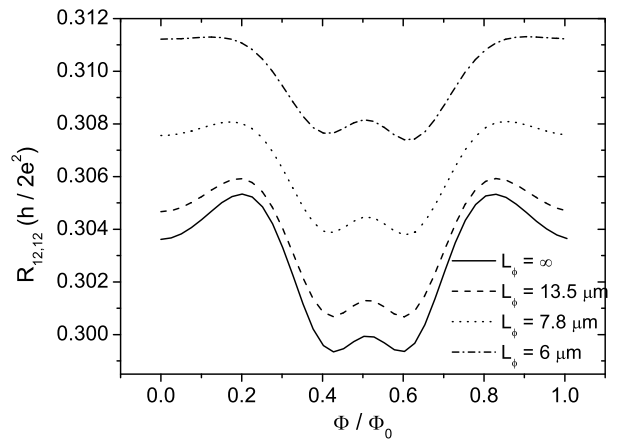}
\caption{ \label{Fig3} Longitudinal resistance of a 4-probe ring
for different values of the phase coherence length.}
\end{figure}

The Hall resistivity is the anti-symmetric part of the resistivity
tensor: $\rho_H = 1/2 \, (\rho_{xy} - \rho_{yx})$. In our case
this comes down to calculating the Hall resistance $R_H = 1/2 \,
(R_{12,34} - R_{34,12})$, which is the transverse resistance part
that is anti-symmetric with respect to the magnetic
flux~\cite{Buettiker2}. Results are shown in Fig.~\ref{Fig2}. The
Hall resistance varies periodically with the magnetic flux through
the ring, the period being the fundamental flux quantum $\Phi_0$.
In the figures, only one fundamental period is shown. It should be
mentioned again that there is no flux through the ring arms, so
the flux is fully contained within the ring. The interesting
effect is that for nonzero flux, a Hall resistance can be
measured, which is not due to a Lorentz force, but which
originates from interference between different trajectories along
the ring like explained previously.

Phase randomizing processes are included in the way explained in
the previous section. The phase coherence length of the sample is
given by $L_\varphi = \sqrt{D \tau_\varphi}$, where $D = 1/2 v_F
l_m$ is the diffusion constant, and $\tau_\varphi = \hbar / 2
\eta$ is the phase relaxation time. The parameter $\eta$
corresponds to the hopping parameter between the attached voltage
probes and the sample. For our system, we varied the phase
coherence length between $6 \mu \text{m}$ and infinity, which is
clearly in the range of validity of our approximation
($\text{system size} > L_\varphi$). When the phase coherence
length is reduced (dashed and dotted lines in the
Fig.~\ref{Fig2}), one can see that the amplitude of the
oscillations decreases. This was to be expected because the Hall
effect we observe is completely due to quantum interference.

Turning to Fig.~\ref{Fig3}, the normal AB oscillations are
observed in the longitudinal resistance. Furthermore, one sees
that the influence of inelastic scattering on the longitudinal
resistance is twofold. Firstly, the mean value of the resistance
will increase by decreasing the phase coherence length because we
are introducing extra scatterers. Second, the amplitude of the AB
oscillations decreases with decreasing coherence length, because
these oscillations are the result of interference effects.
\begin{figure}
\includegraphics[width = 8cm]{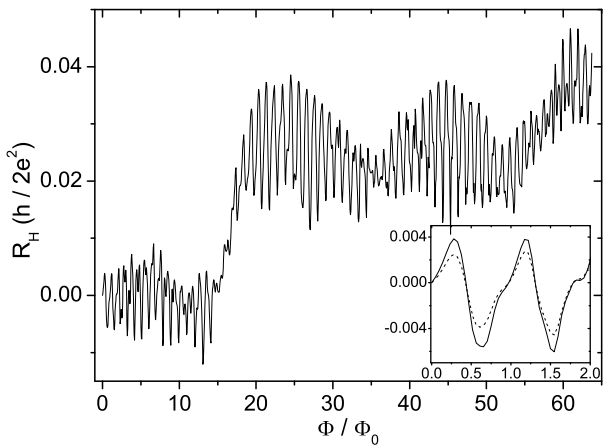}
\caption{ \label{Fig4} Hall resistance $R_H$ of the 4-probe ring
when a magnetic field is applied across the whole sample. $\Phi$
is defined as the flux through the mean radius of the ring. The
effect of inelastic scattering is shown in the inset, for the
first 2 oscillations in the resistance: $L_\varphi = \infty$
(solid line) and $L_\varphi = 7.8 \mu \text{m}$ (dashed).}
\end{figure}

When comparing the magnitude of the Hall resistance in
Fig.~\ref{Fig2} with the longitudinal resistance shown in
Fig.~\ref{Fig3}, we see that the oscillations have the same order
of magnitude. Since normal Aharonov-Bohm (AB) oscillations can be
observed experimentally quite easily nowadays, this means that it
is also feasible to measure the Hall effect in the 4-probe ring.

However, in an experimental setup it is difficult to confine the
magnetic flux to a region inside the ring, and therefore the
magnetic field is applied across the whole sample. In
Fig.~\ref{Fig4}, we show calculation results for this case. The
Hall resistance $R_H$ is not anymore strictly periodic with
respect to the flux. Nevertheless quasi-periodic oscillations are
visible resulting from quantum interference, whose amplitude will
decrease when introducing inelastic scattering (see the inset of
Fig.~\ref{Fig4}). A beating pattern can be observed in the
oscillations, which is a result of having several channels open
for conduction in the ring arms; trajectories for different
channels surround slightly different areas and thus fluxes.
Compared to Fig.~\ref{Fig2}, the Hall resistance also gets a
non-zero offset: this contribution is a result of the Lorentz
force acting on the electrons when there is a flux present in the
ring arms.

\section{Conclusion}
Incorporating the effect of phase randomizing processes in a
phenomenological way can be done with the attachment of extra
voltage probes to sample, an idea originally proposed by
B\"{u}ttiker~\cite{Buettiker1}. In this paper, we have described a
method for treating inelastic effects based on this idea, but such
that a recursive technique for calculating the transmission
coefficients can still be used. Doing so, one is able to treat the
problem in a numerically very efficient way, which was not
possible within the original proposal. The approach however
consists of neglecting multiple inelastic scattering and is
therefore only valid in a regime where the phase coherence length
is larger than the system size.

The method was applied to an experimentally realizable ring with
four attached contacts, and a Hall effect was observed which is
due to quantum interference rather than an implicit Lorentz force
acting on the electrons. This Hall effect disappears with
decreasing the phase coherence length.

\end{document}